\documentclass[11pt]{article}
\usepackage{amsmath}
\usepackage{amssymb}
\usepackage{bbm}
\usepackage{epsfig}
\allowdisplaybreaks[4]
\begin{document}
\title{Neutral minima in two-Higgs doublet models}
\author{A. Barroso$^{1,}$~\footnote{barroso@cii.fc.ul.pt},
P.M. Ferreira$^{1,}$~\footnote{ferreira@cii.fc.ul.pt} and
R. Santos$^{1,2,}$~\footnote{rsantos@cii.fc.ul.pt}\\
$^1$ Centro de F\'{\i}sica Te\'orica e Computacional, Faculdade de Ci\^encias,\\
Universidade de Lisboa, Av. Prof. Gama Pinto, 2, 1649-003 Lisboa, Portugal
\\
$^{2}$ Department of Physics, Royal Holloway, University of London,
\\ Egham, Surrey TW20 0EX United Kingdom }
\date{February, 2007}
\maketitle \noindent

{\bf Abstract.} We study the neutral minima of two-Higgs doublet
models, showing that these potentials can have at least two such
minima with different depths. We analyse the phenomenology of these
minima for the several types of two-Higgs doublet potentials, where
CP is explicitly broken, spontaneously broken or preserved. We
discover that it is possible to have a neutral minimum in these
potentials where the masses of the known particles have their
standard values, with another deeper minimum where those same
particles acquire different masses.

\vspace{1cm}

\section{Introduction}

The standard model (SM) of electroweak interactions is a remarkably
successful theory, but its scalar sector is yet untested. Numerous
theories with a larger scalar content have been proposed over the
years in the attempt to increase the predictive power of the model
and offer explanations for problems such as baryogenesis, CP
violation, the hierarchy question and others. In models with more
than one scalar, the possibility of the scalar potential having more
than one minimum arises, and some of those minima can break the
$SU(3)_C \times U(1)_{em}$ gauge symmetry of the SM. Thus one is
left with the possibility of imposing charge and/or colour breaking
bounds on the parameters of the theory: excluding those combinations
of parameters for which the potential's deeper minimum breaks charge
and/or colour conservation. This appealing idea was first considered
in the framework of supersymmetric theories~\cite{fre} and applied
extensively to such models~\cite{cbss}. Recently, it has also been
applied to the Zee model~\cite{zee} and to the two-Higgs doublet
model (2HDM)~\cite{nos, cruz}. The results obtained in this last
reference were generalized in ref.~\cite{nhdm} to models with an
arbitrary number of Higgs doublets~\cite{nhdm2}.

The results of~\cite{nos} may be summarised as follows: in the 2HDM,
whenever a minimum that preserves charge and CP (we dub these
``normal" minima) exists, that minimum is certainly deeper than any
charge breaking (CB) stationary point. Further, the global minimum
of the potential is a normal one, and the CB stationary point ends
up being a saddle point. A similar conclusion holds for the
spontaneous breaking of the CP symmetry: when a normal minimum
exists, it is certainly deeper than any possible CP stationary
point, and the global minimum is a normal one. However, unlike the
CB case, the question of whether the CP stationary point can be a
minimum was left unanswered in ref.~\cite{nos}. There is then the
possibility that the potential may have a CP minimum, with a normal
one lying below it.

The importance of these results is to ensure the stability of the
normal minimum against spontaneous charge or CP breaking: if one
finds a normal minimum in the 2HDM, one may rest assured that there
is no deeper CB or CP minimum for which the system may eventually
evolve via tunneling. However, the work of ref.~\cite{nos} did not
answer the following question: how many normal minima does the 2HDM
have? Are they all acceptable minima, with phenomenology according
to the experimental data? And if there are several normal minima,
which is the deepest one? Several of these questions were addressed
for the 2HDM in recent papers~\cite{mani}, but in this paper we will
focus on other aspects not treated in those works. A recent
work~\cite{mann} studied the minimum structure of the
next-to-minimal Supersymmetric Standard Model.

In the present work we will take a close look at the normal minimum
structure of several types of 2HDM potentials. We will show that
there is indeed the possibility that several normal minima coexist
in the 2HDM, and if they are not degenerate, they will give rise to
different phenomenologies. This paper is structured as follows: in
section~\ref{sec:int} we will introduce our formalism, write down
the most general 2HDM potential and analyse the several types of
theories one obtains by imposing symmetries on the model. In
section~\ref{sec:vac} we will discuss the vacuum structure of the
model and review the results of~\cite{nos} about the differences in
depths of the scalar potential at two of these possible vacua. In
section~\ref{sec:cpv} we will analyse normal minima in potentials
where CP is broken - explicitly or spontaneously - and, in
section~\ref{sec:cpc}, in potentials where CP is always left
unbroken. We conclude, in section~\ref{sec:conc}, with a general
overview.

\section{The 2HDM potentials}
\label{sec:int}

Let us consider two $SU(2)$ doublets, with hypercharge $Y = 1$,
given by
\begin{equation}
\Phi_1 = \begin{pmatrix} \varphi_1 + i \varphi_2 \\ \varphi_5 + i
\varphi_7
\end{pmatrix} \;\; , \;\; \Phi_2 = \begin{pmatrix} \varphi_3 + i \varphi_4 \\
\varphi_6 + i \varphi_8 \end{pmatrix} \;\; .
\end{equation}
where all the $\varphi_i$ are real functions. Their numbering may
seem odd, but it simplifies the writing of the scalar mass matrices.
With these two fields one can build four $SU(2) \times U(1)$ real
quadratic invariants,
\begin{align}
x_1\,\equiv\,|\Phi_1|^2 &= \varphi_1^2 + \varphi_2^2 + \varphi_5^2 +
\varphi_7^2
\nonumber \\
x_2\,\equiv\,|\Phi_2|^2 &= \varphi_3^2 + \varphi_4^2 + \varphi_6^2 +
\varphi_8^2
\nonumber \\
x_3\,\equiv\,Re(\Phi_1^\dagger\Phi_2) &= \varphi_1 \varphi_3 +
\varphi_2
\varphi_4 + \varphi_5\varphi_6 + \varphi_7\varphi_8 \nonumber \\
x_4 \,\equiv\,Im(\Phi_1^\dagger\Phi_2) &= \varphi_1 \varphi_4 -
\varphi_2 \varphi_3 + \varphi_5 \varphi_8 - \varphi_6 \varphi_7
\;\;.
\label{eq:x}
\end{align}
The most general 2HDM potential (for an overview, see for
instance~\cite{sher, ginz}) is therefore a polynomial on the $x$'s,
with all possible linear and quadratic terms in these variables.
That is,
\begin{align}
V \;\;=& \;\; a_1\, x_1\, + \,a_2\, x_2\, + \,a_3 x_3 \,+\, a_4 x_4
\,+\, b_{11} \, x_1^2\, +\, b_{22}\, x_2^2\, +\, b_{33}\, x_3^2\,
+\, b_{44}\, x_4^2\, +\,
\nonumber \\
 & \;\; b_{12}\, x_1 x_2\, +\, b_{13}\, x_1 x_3\, + b_{14}\,
x_1 x_4\, +\,b_{23}\, x_2 x_3 +\,b_{24}\, x_2 x_4 +\,b_{34}\, x_3
x_4\;\; , \label{eq:pot}
\end{align}
where the coefficients $a_i$ and $b_{ij}$ are all real, the former
having dimensions of mass squared and the latter being
dimensionless. Under a CP transformation of the form $\Phi_i
\rightarrow \Phi_i^*$, $x_1$, $x_2$ and $x_3$ remain the same but
$x_4$ switches signal. Thus, the terms of the potential which are
linear in $x_4$ break CP explicitly. The most general explicit
CP-breaking potential has therefore 14 real parameters. Through an
appropriate choice of basis for $\{\Phi_1\,,\,\Phi_2\}$ it is
possible to reduce that number to 11~\cite{hab}. It is easy to see
how to reduce the number of parameters to 12. The quadratic part
of~\eqref{eq:pot} may be written as
\begin{equation}
a_1\, x_1\, + \,a_2\, x_2\, + \,a_3 x_3 \,+\, a_4 x_4 \;=\;
\begin{bmatrix} \Phi_1^\dagger & \Phi_2^\dagger \end{bmatrix} \,
\begin{bmatrix} a_1 & \displaystyle{\frac{a_3\,-\,i\,a_4}{2}} \\
\displaystyle{\frac{a_3\,+\,i\,a_4}{2}} & a_2
\end{bmatrix} \,\begin{bmatrix} \Phi_1 \\ \Phi_2
\end{bmatrix} \;\;\; .
\label{eq:amat}
\end{equation}
Through an unitary transformation on the fields, it is possible to
diagonalize this $2\times 2$ matrix. Written in terms of the new
fields, the quadratic terms are now just two, whereas the quartic
potential continues to have 10 parameters. In short, given the most
general 2HDM potential, it is always possible to choose a field
basis for which one has $a_3 = a_4 = 0$. To ensure that a minimum
exists away from the origin, then at least one of the parameters
$a_1$, $a_2$, has to be negative. Equation~\eqref{eq:tan4} below
shows us that one of the parameters
$\{b_{14}\,,\,b_{24}\,,\,b_{34}\}$ may be expressed in terms of the
two others, which leaves us with 11 independent real parameters.

If we wish to study a potential for which CP is not explicitly
broken, we must set to zero all terms linear in $x_4$ in
eq.~\eqref{eq:pot}: $a_4 = b_{14} = b_{24} = b_{34} = 0$. Again, we
may choose a basis in which $a_3 = 0$, and we are left with a
9-parameter potential. For this potential, one has the possibility
of spontaneous breaking of CP.

Finally, we may impose further symmetries on the model such that
spontaneous CP breaking becomes impossible, and one is left with
potentials containing seven independent real parameters. For
instance, if one asks that the potential is invariant under a global
$Z_2$ symmetry ($\Phi_1 \rightarrow -\Phi_1$, $\Phi_2 \rightarrow
\Phi_2$), then the terms linear in $x_3$ and $x_4$ vanish and the
potential becomes
\begin{align}
V_A \;\;=& \;\; a_1\, x_1\, + \,a_2\, x_2\,+\, b_{11} \, x_1^2\, +\,
b_{22}\, x_2^2\, +\, b_{33}\, x_3^2\, +\, b_{44}\, x_4^2\, +\,
b_{12}\, x_1 x_2 \;\;\; .\label{eq:potA}
\end{align}
We call this the potential A. It is very important to have $b_{33}
\neq b_{44}$ in this model, otherwise one has a massless
axion~\footnote{In fact, without this restriction, the potential has
the $U(1)$ symmetry of model B without its soft breaking term.}.
Alternatively, with a $U(1)$ symmetry ($\Phi_1 \rightarrow
e^{i\alpha} \Phi_1$, $\Phi_2 \rightarrow \Phi_2$), the model is also
greatly simplified and one obtains the potential B,
\begin{align}
V_B \;\;=& \;\; a_1\, x_1\, + \,a_2\, x_2\, + \,a_3\, x_3\, +\,
b_{11} \, x_1^2\, +\, b_{22}\, x_2^2\, +\, b_{33}\, (x_3^2\, +\,
x_4^2)\, +\, b_{12}\, x_1 x_2 \;\;\; ,\label{eq:potB}
\end{align}
where the symmetry imposes $b_{33}\,=\,b_{44}$. This potential is
the analogous of the supersymmetric one. Notice that the $a_3$ term
softly breaks the $U(1)$ symmetry. It was left there to prevent the
appearance of a massless axion in the theory. Unlike the most
general case, here one cannot ``rotate away" the $a_3$ term without
introducing further parameters in the quartic terms of $V_B$. In
order to ensure that a minimum exists away from the origin, a
trivial calculation shows that we must have $a_3^2\,>\,4\,a_1\,a_2$.
As was shown in ref.~\cite{vel}, the imposition of these symmetries
makes it impossible for spontaneous CP breaking to occur in models A
or B. Another property of these symmetries is that they are exactly
those that are necessary to prevent flavour changing neutral
currents (FCNC) in the theory, a phenomenon that plagues both the
potential~\eqref{eq:pot} and its restriction to the case of CP not
being explicitly broken.

Another possible symmetry to impose on the potential is that it
remains invariant under an interchange between both fields ($\Phi_1
\leftrightarrow \Phi_2$). Then the potential becomes~\cite{bran}
\begin{align}
V_C \;\;=& \;\; a_1\,( x_1\, + \, x_2)\, + \,a_3\, x_3\, + \, b_{11}
\, (x_1^2\, +\, x_2^2)\, +\, b_{33}\, x_3^2\, +\, b_{44}\, x_4^2 \,+
\nonumber \\
 & \;\; b_{12}\, x_1 x_2 \, +\, b_{13}\,(x_1\, + \,
x_2)\,x_3 \;\;\; .\label{eq:potBr}
\end{align}
Again, no basis changes can reduce the number of parameters in this
model.

The potential $V_C$ was first studied by Branco and Rebelo
in~\cite{bran}, introducing an extra term of the form $a_2\,(x_1\,
-\, x_2)$ to softly break the symmetry imposed on the $\Phi$ fields.
The consequence is that the resulting potential may have spontaneous
CP breaking, without FCNC. Similarly, adding to $V_A$ a term of the
form  $a_3\,x_3$, one breaks softly the $Z_2$ symmetry that
characterizes that potential. Once again, this gives rise to the
possibility of spontaneous CP breaking. In fact, in ref.~\cite{hab}
it was shown that the potentials $A$ and $C$ are equivalent - it is
possible to obtain the one from the other with a basis
transformation.

Interestingly, no soft breaking of the $U(1)$ symmetry of the
potential $V_B$ gives rise to spontaneous CP breaking. The potential
written in eq.~\eqref{eq:potB} has already a soft breaking term -
the $a_3$ term - and the only remaining term quadratic in the fields
one could add to the potential would be $a_4\,x_4$ - and that term
explicitly breaks the CP symmetry.

Let us summarise what we have discussed about the 2HDM potentials in
terms of the number of independent real parameters that they have:
\begin{itemize}
\item Potentials where CP is explicitly broken, which have 11 real parameters. FCNCs
occur.
\item Potentials with 9 real parameters where explicit CP conservation has been imposed, but for which one
might have spontaneous CP breaking. Again, FCNCs occur.
\item Potentials with 7 real parameters where, besides explicit CP conservation,
one has also imposed a discrete symmetry ($Z_2$ or permutation of
$\Phi_1$ and $\Phi_2$). No spontaneous symmetry breaking occurs,
FCNCs are excluded. With an additional term that breaks softly the
discrete symmetry, spontaneous CP might arise.
\item Potentials with 6 real parameters, with explicit CP conservation and invariant under a global
$U(1)$ symmetry. Again there is no FCNC but these models have a
massless axion. If one adds to the potential a term that softly
breaks the global $U(1)$, that axion acquires a mass but there is no
possibility of spontaneous CP breaking.
\end{itemize}
There is an interesting possible classification of these potentials,
which we will present in appendix~\ref{ap:cla}.

\section{The stationary points of the 2HDM}
\label{sec:vac}

The vacuum structure of the 2HDM is very simple~\cite{msher}, with
three different types of vev configuration. For a charge and CP
conserving vacuum, the fields $\Phi_1$ and $\Phi_2$ have vevs of the
form
\begin{equation}
\Phi_1 \,\rightarrow\,\begin{pmatrix} 0 \\ v_1
\end{pmatrix} \;\;\; , \;\;\;\Phi_2 \,\rightarrow\,\begin{pmatrix} 0 \\
v_2 \end{pmatrix} \label{eq:vevn}
\end{equation}
where $v_1$ and $v_2$ are real numbers. For a vacuum that breaks
charge conservation, we have
\begin{equation}
\Phi_1 \,\rightarrow\,\begin{pmatrix} 0 \\ v^\prime_1
\end{pmatrix} \;\;\; , \;\;\;\Phi_2 \,\rightarrow\,\begin{pmatrix} \alpha \\
v^\prime_2 \end{pmatrix}\;\;\; . \label{eq:vevcb}
\end{equation}
Again, all vevs in this equation are real. Finally, a CP breaking
vacuum occurs when the fields acquire vevs of the form
\begin{equation}
\Phi_1 \,\rightarrow\,\begin{pmatrix} 0 \\
v^{\prime\prime}_1\,+\,i\delta
\end{pmatrix} \;\;\; , \;\;\;\Phi_2 \,\rightarrow\,\begin{pmatrix} 0 \\
v^{\prime\prime}_2 \end{pmatrix}\;\;\; . \label{eq:vevcp}
\end{equation}
Vacua with $\alpha$ and $\delta$ simultaneously non-zero are not
considered because the minimisation conditions of the potential
forbid them~\footnote{Except for a very special case in the explicit
CP breaking potential. Even in that case, though, via a basis
change, that vacuum may be reduced to one with $\alpha\,\neq\,0$ and
$\delta\,=\,0$.}.

The structure of the potential of eq.~\eqref{eq:pot} is such that it
may have several stationary points, and they may be of different
natures. In ref.~\cite{nos} we obtained a remarkable result that
relates the difference in the depths of the potential in each of the
three types of possible stationary points. Namely, the difference
between the value of the potential at a CB stationary point,
$V_{CB}$, and the value of potential at a normal stationary point,
$V_N$, is given by
\begin{equation}
V_{CB}\;-\;V_N\;\; = \;\; \frac{M^2_{H^\pm}}{2\,v^2} \;\left[
(v^\prime_1\,v_2 \;-\;v^\prime_2\,v_1)^2\; + \;
\alpha^2\,v_1^2\right]\;\;\; , \label{eq:difcb}
\end{equation}
where $v^2\,=\,v_1^2\,+\,v_2^2$ and $M^2_{H^\pm}$ is the value of
the squared mass of the charged Higgs scalar, evaluated at the
normal stationary point. What this equation tells us is that {\em
when} the normal stationary point is a minimum~\footnote{Even more
generally, when it is a stationary point for which $M^2_{H^\pm}>0$.}
then, all of the eigenvalues of the squared scalar mass matrices
being positive, we will have $M^2_{H^\pm}>0$ and the deeper
stationary point will be the normal minimum. Furthermore, in
ref.~\cite{nos} we were able to prove that the CB stationary point
is unique, and that eq.~\eqref{eq:difcb}, when $M^2_{H^\pm}>0$, also
implies that it is, necessarily, a saddle point. The stability of
the normal minimum against tunneling to a deeper charge breaking
stationary point is thus ensured in the 2HDM.

A similar result holds for the comparison between the CP and normal
stationary points. In the case where it makes sense to define a CP
symmetry - that is, in potentials where it is not explicitly broken
- the difference between the value of the potential in the
stationary points with vevs given by eqs.~\eqref{eq:vevn}
and~\eqref{eq:vevcp} is given by~\cite{nos}
\begin{equation}
V_{CP}\;-\;V_N\;\; = \;\; \frac{M^2_A}{2\,v^2} \;\left[
(v^{\prime\prime}_1\,v_2 \;-\;v^{\prime\prime}_2\,v_1)^2\; + \;
\delta^2\,v_2^2\right]\;\;\; . \label{eq:difcp}
\end{equation}
$M^2_A$ is now the value of the squared pseudoscalar mass at the
normal stationary point. Again, if this squared mass is positive,
for instance if the normal stationary point is a minimum, then the
normal minimum is indeed deeper than the CP breaking one. Again, the
CP stationary point being unique, the stability of the normal
minimum against tunneling is guaranteed. However, unlike the CB
case, we were not able to prove, in ref.~\cite{nos}, that the CP
stationary point is necessarily a saddle point when the normal
stationary point is a minimum. Thus, even though we have proved that
the global minimum is a normal one, we cannot discard the
possibility that above it a CP breaking minimum exists.

The final case of interest occurs in the potential with explicit CP
breaking. In that situation, both the vevs of eq.~\eqref{eq:vevn}
and those of eq.~\eqref{eq:vevcp} have the same physical meaning:
both stationary points break the same symmetry and the existence of
a relative phase between the vevs of each field does not distinguish
them. For this potential, we refer to the vevs of~\eqref{eq:vevn} as
the ``$N_1$ stationary point", and those of~\eqref{eq:vevcp} as the
``$N_2$ stationary point". It may be shown~\cite{hab} that it is
possible to pass from a vev structure of the type $N_2$ to vevs of
the $N_1$ type by a specific basis transformation - the complex
phase of eq.~\eqref{eq:vevcp} is absorbed in the parameters of the
potential, and the final vevs of both $\Phi_1$ and $\Phi_2$ are
real. However, this {\em does not} mean that the $N_1$ and $N_2$
stationary points are the same - simply that there is a field basis
for which $N_2$ may be written with real vevs. In that new basis,
however, the $N_1$ vevs would have acquired a relative complex
phase. In ref.~\cite{nos} we found an interesting relationship
between the difference in the depths of the potential at the $N_1$,
$N_2$ stationary points. Namely,
\begin{equation}
V_{N_2}\;-\;V_{N_1} \;\; = \;\;\frac{1}{2}\left[
\left(\frac{M^2_{H^\pm}}{v^2}\right)_{N_1}\;-\;\left(\frac{M^2_{H^\pm}}{v^2}
\right)_{N_2} \right] \;\left[(v^{\prime\prime}_1\,v_2 \;-\;
v^{\prime\prime}_2\,v_1)^2\; + \; \delta^2\,v_2^2\right]\;\;\; .
\label{eq:difn12}
\end{equation}
In this equation we have $(v^2)_{N_1}\,=\,v_1^2\,+\,v_2^2$ and
$(v^2)_{N_2}\,=\,{v^{\prime\prime}_1}^2\,+\,{v^{\prime\prime}_2}^2\,+\,\delta^2$,
and $(M^2_{H^\pm})_{N_{1,2}}$ are the squared charged scalar masses
at each of the $N_1$, $N_2$ stationary points. A similar expression
was found in ref.~\cite{mani}. Equation~\eqref{eq:difn12} tells us
the deepest stationary point will be the one with the largest ratio
between the square of the charged Higgs mass and the $v^2$.

Let us now consider the stationarity equations that give rise to the
different stationary points that we have been discussing. We have
mentioned that the CB and CP stationary points are unique since they
are given by linear equations on the vevs. However, this is not true
for the normal stationary point. Let us begin with the most general
2HDM potential written in a basis where $a_3\,=\,a_4\,=\,0$. We make
$\varphi_5\,=\,v_1$ and $\varphi_6\,=\,v_2$ in eq.~\eqref{eq:x} and
minimise~\eqref{eq:pot} by solving the equations $\partial
V/\partial \varphi_i\,=\,0$, $i\,=\,1\ldots 8$. Most of these
equations are trivially satisfied. The non-trivial ones are
\begin{align}
\frac{\partial V}{\partial \varphi_5} & =\;\; 2\,a_1\,v_1 \,+\,
4\,b_{11}\,v_1^3 \,+\, 2\,(b_{12} \,+\, b_{33})\,v_1\,v_2^2 \,+\,
3\,b_{13}\,v_1^2\,v_2 \,+\, b_{23}\,v_2^3\;=\;0
\nonumber \\
\frac{\partial V}{\partial \varphi_6} & =\;\; 2\,a_2\,v_2 \,+\,
4\,b_{22}\,v_2^3 \,+\, 2\,(b_{12} \,+\, b_{33})\,v_1^2\,v_2 \,+\,
b_{13}\,v_1^3 \,+\, 3\,b_{23}\,v_1\,v_2^2\;=\;0 \;\;\;
,\label{eq:sta}
\end{align}
and
\begin{align}
\frac{\partial V}{\partial \varphi_7} & =\;\; v_1\,\left(
b_{14}\,v_1^2\,+\,b_{24}\,v_2^2\,+\,b_{34}\,v_1\,v_2\right) &\;=\,0
\;\;\; \;
\nonumber \\
\frac{\partial V}{\partial \varphi_8} & =\;\; -\,v_2\,\left(
b_{14}\,v_1^2\,+\,b_{24}\,v_2^2\,+\,b_{34}\,v_1\,v_2\right) &\;=\,0
\;\;\; . \label{eq:stacp}
\end{align}
Notice that one cannot have solutions of the form
$\{v_1\,=\,0\,,\,v_2\,\neq\,0\}$ or
$\{v_1\,\neq\,0\,,\,v_2\,=\,0\}$, unless some parameters of the
potential are set to zero ($b_{23}$, $b_{24}$ and $b_{13}$, $b_{14}$
respectively). Since there is no symmetry forcing those parameters
to be zero, they have to be present in the potential. We now define
the usual polar coordinates $v_1\,=\,v\,\cos\beta$ and
$v_2\,=\,v\,\sin\beta$. A trivial solution of these equations is
clearly $v\,=\,0$. Excluding that case, the stationarity
conditions~\eqref{eq:sta} become
\begin{equation}
v^2\;=\; -\,\frac{1}{\cos^2\beta}\;\frac{2\,a_1}{b_{23}\,\tan^3\beta
\,+\, 2\,(b_{12} \,+\, b_{33})\,\tan^2\beta \,+\,
3\,b_{13}\,\tan\beta \,+\, 4\,b_{11}} \label{eq:v2}
\end{equation}
and
\begin{align}
 & - a_2\,b_{23}\,\tan^4\beta\,+\,\left[4\,a_1\,b_{22} -
2\,a_2\,(b_{12} + b_{33})\right]\,\tan^3\beta \nonumber \\
 & +
3\,(a_1\,b_{23} - a_2\,b_{13})\,\tan^2\beta
\,+\,\left[2\,a_1\,(b_{12} + b_{33}) - 4\,a_2\,b_{11}
\right]\,\tan\beta \, +\, a_1\,b_{13} \;=\;0 \label{eq:tan}
\end{align}
and both equations~\eqref{eq:stacp} reduce to
\begin{equation}
b_{24}\,\tan^2\beta\,+\,b_{34}\,\tan\beta\,+\,b_{14}\;=\;0\;\;\;.
\label{eq:tan4}
\end{equation}
Eq.~\eqref{eq:v2} tells us that, other than its sign, the value of
$v$ is determined unequivocally by $\tan\beta$. Eq.~\eqref{eq:tan}
is a quartic equation on $\tan\beta$, having at most four possible
real solutions. These two equations describe therefore {\em eight}
possible solutions $\{v_1\,,\,v_2\}$, due to the ambiguity on the
sign of $v$. The 2HDM potential~\eqref{eq:pot} is however invariant
under the transformation $\Phi_1\rightarrow -\Phi_1$ and
$\Phi_2\rightarrow -\Phi_2$, so that these eight solutions
correspond to only {\em four} different physical scenarios. Adding
the trivial solution $v_1\,=\,v_2\,=\,0$, we have a total of nine
solutions. However, we must contend with eq.~\eqref{eq:tan4} as
well, which is a quadratic equation on $\tan\beta$. Then, there are
at most {\em two} different values of $\tan\beta$ which satisfy all
equations. This means that we have a maximum of {\em five}
stationary points.

For potentials with explicit CP conservation,
equations~\eqref{eq:stacp} are trivially satisfied, since
$b_{14}\,=\,b_{24}\,=\,b_{34}\,=\,0$. Therefore,
equation~\eqref{eq:tan4} does not exist and the potential could have
a total of {\em nine} stationary points.

At this point we ask: can we have more than one normal minimum, with
different depths? The answer is yes, and to see this we make use of
Morse's inequalities~\cite{morse}: for a given real function of two
variables, let $m_0$, $m_1$ and $m_2$ be the number of its minima,
saddle points and maxima, respectively. For a polynomial function in
$v_1$ and $v_2$, bounded from below, such as the one we are dealing
with, Morse's inequalities state that:
\begin{itemize}
\item $m_0\,\geq\,1$;
\item $m_1\,\geq\,m_0\,-\,1$;
\item $m_0\,-\,m_1\,+\,m_2\,=\,1$.
\end{itemize}
We know that the 2HDM potential has $m_0\,+\,m_1\,+\,m_2\,=\,2\,n +
1$ stationary solutions, $n\,=\,0,\,\ldots,\,4$: at most $2\,n$ real
roots of eqs.~\eqref{eq:v2},~\eqref{eq:tan} and~\eqref{eq:tan4} plus
the trivial solution $v_1\,=\,v_2\,=\,0$. Hence we find that
$m_0\,+\,m_2\,=\,n\,+\,1$. Let us analyse the several possibilities
for the number of minima $m_0$, depending on the number of solutions
$n$. Simply counting all the different combinations of extrema leads
us to:
\begin{itemize}
\item $n\,=\,0$: we have necessarily $m_0\,=\,1$, the minimum is
unique but is located at the origin, $v_1\,=\,v_2\,=\,0$, which
means that there is no $SU(2)_W\times U(1)_Y$ symmetry breaking.
This case is excluded on physical grounds.
\item $n\,=\,1$: we find two possibilities:
\begin{itemize}
\item $m_0\,=\,1$, which is the previous case.
\item $m_0\,=\,2$, which means two degenerate minima away from the
origin, related to one another by a change of sign of the vevs. This
situation corresponds to an acceptable symmetry breaking and it
means that there are no normal minima with different depths. This
would be the ``standard" situation.
\end{itemize}
\item $n\,=\,2$: there are three possibilities:
\begin{itemize}
\item $m_0\,=\,1$ or $2$ are like the previous cases.
\item $m_0\,=\,3$, one uninteresting minimum at the origin, and two
degenerate ones away from it. This situation would also be the
``standard" one, as there would be no normal minima with different
depths.
\end{itemize}
\item $n\,=\,3$: there are now three qualitatively different cases:
\begin{itemize}
\item $m_0\,=\,1\,,\,2\,,\,3$, like above.
\item $m_0\,=\,4$, this case corresponds to {\em two pairs of degenerate
minima} away from the origin. Nothing forces these two pairs of
minima to have the same depth. We might therefore have one normal
minimum deeper than another.
\end{itemize}
\item $n\,=\,4$: we have:
\begin{itemize}
\item $m_0\,=\,1\,,\,2\,,\,3\,,\,4$, like above.
\item $m_0\,=\,5$ is similar to the $m_0\,=\,4$ case examined above, with an
extra minimum present at the origin.
\end{itemize}
\end{itemize}
This trivial analysis shows us that, if there are more than two
solutions for $\tan\beta$, then the 2HDM may have more than one
normal minimum away from the origin at different depths. However, no
more than two such minima can exist.

Why is this interesting? We already know that in the 2HDM, when a
normal minimum exists, then the {\em global} minimum of the theory
is normal~\cite{nos}. However, another interesting possibility might
arise: the minimisation equations may give us a normal minimum $N_1$
with vevs $\{v_1\,,\,v_2\}$ and another normal minimum $N_2$ with
vevs $\{\hat{v}_1\,,\,\hat{v}_2\}$. Which is the deepest? Suppose
that we find $N_1$ for which the vevs are such that the SM
phenomenology is satisfied, namely,
$v_1^2\,+\,v_2^2\,=\,(246\;\mbox{GeV})^2$ and the $W$ and $Z$ boson
masses are according to their experimental values; but $N_2$ is
deeper, for which the sum of the squared vevs
$\hat{v}_1^2\,+\,\hat{v}_2^2$ has a completely different value,
contrary to experimental data. This situation is clearly undesirable
since it can lead to tunneling between an acceptable minimum and one
with gauge boson and fermion masses different from their measured
values. A trivial calculation (see appendix~\ref{ap:dem}) shows us
that the difference in depth of the potential at $N_1$ and $N_2$ is
given by
\begin{equation}
V_{N_2}\;-\;V_{N_1} \;\; = \;\;\frac{1}{2}\left[
\left(\frac{M^2_{H^\pm}}{v^2}\right)_{N_1}\;-\;\left(\frac{M^2_{H^\pm}}{v^2}
\right)_{N_2} \right] \;\left(v_1\,\hat{v}_2 \;-\;
v_2\,\hat{v}_1\right)^2\;\;\; , \label{eq:difnn}
\end{equation}
a result very similar to eq.~\eqref{eq:difn12}. Again, it is the
squared mass of the charged scalars  divided by the respective $v^2$
which ``controls" the difference in depths of the potential at the
stationary points. As we will shortly show, it is possible,
depending on the value of the parameters of the potential, to have
such ``coexisting" minima in situations of interest for particle
physics phenomenology.

\section{Neutral minima in potentials with CP breaking}
\label{sec:cpv}

As mentioned before, one of the most interesting features of the
2HDM is that it allows for breaking of the CP symmetry. We will now
analyse the possibility of coexistence of neutral minima in this
potential, for the two cases of CP breaking: explicit and
spontaneous.

\subsection{Potential with explicit CP breaking}
\label{sec:expcp}

As we saw earlier, for the 2HDM potential with explicit CP breaking
the value of $\tan\beta$ is determined by a set of two
equations,~\eqref{eq:tan} and~\eqref{eq:tan4}, one quartic and
another quadratic. This means that we may have, at most, two
different values of $\tan\beta$. We are therefore in the $n\,=\,2$
case discussed above. As was shown, this means that the normal
minimum exists and is unique.

This conclusion, however, was derived for stationary points with
real vevs, of the form~\eqref{eq:vevn}. We know that in this
potential CP is not defined and as such stationary points of the
form of eq.~\eqref{eq:vevcp}, in which the vevs have a relative
complex phase, have the same physical relevance as stationary points
with real vevs. Equation~\eqref{eq:difn12} gives us the difference
in depth of the 2HDM potential with explicit CP breaking at two
neutral stationary points, $N_1$ (with real vevs) and $N_2$ (with
vevs with a relative complex phase). This result was first obtained
in ref.~\cite{nos}, but one question was left open in that work: is
it {\em possible} to find values of the potential such that one can
find two minima that verify eq.~\eqref{eq:difn12}? We will perform a
numerical study to prove that such a situation is indeed possible.

The stationarity conditions for a $N_1$ stationary point with vevs
such as those of eq.~\eqref{eq:vevn} were shown in
eq.~\eqref{eq:sta}. Similarly, for a $N_2$ stationary point, with
vevs of the form~\eqref{eq:vevcp}, the stationarity conditions are
\begin{align}
 & 2\,a_1\,v^{\prime\prime}_1 \,+\,
4\,b_{11}\,v^{\prime\prime}_1\,({v^{\prime\prime}_1}^2\,+\,\delta^2)
\,+\, 2\,(b_{12} \,+\,
b_{33})\,v^{\prime\prime}_1\,{v^{\prime\prime}_2}^2 \,+\,
b_{13}\,(3\,{v^{\prime\prime}_1}^2\,+\,\delta^2)\,v^{\prime\prime}_2
\,+\, b_{23}\,{v^{\prime\prime}_2}^3 & \nonumber \\
 & \,-\,2\,b_{14}\,v^{\prime\prime}_1\,v^{\prime\prime}_2\,\delta \,-\,
b_{34}\,{v^{\prime\prime}_2}^2\,\delta & =\;0 \vspace{0.3cm} \nonumber \\
 & 2\,a_2\,v^{\prime\prime}_2 \,+\, 4\,b_{22}\,{v^{\prime\prime}_2}^3
\,+\,
2\,b_{12}\,v^{\prime\prime}_2\,({v^{\prime\prime}_1}^2\,+\,\delta^2)
\,+\, 2\,b_{33}\,{v^{\prime\prime}_1}^2\,v^{\prime\prime}_2 \,+\,
b_{13}\,v^{\prime\prime}_1\,({v^{\prime\prime}_1}^2\,+\,\delta^2)
 & \nonumber \\
 & \,+\,3\,b_{23}\,v^{\prime\prime}_1\,{v^{\prime\prime}_2}^2
\,-\,b_{14}\,({v^{\prime\prime}_1}^2\,+\,\delta^2)\,\delta
 \,-\,3\,b_{24}\,{v^{\prime\prime}_2}^2\,\delta \,-\,2\,b_{34}\,v^{\prime\prime}_1\,v^{\prime\prime}_2\,\delta
 \,+\,2\,b_{44}\,v^{\prime\prime}_2\,\delta^2 & =\;0 \vspace{0.3cm} \nonumber \\
 & 2\,a_1\,\delta \,+\,
4\,b_{11}\,\delta\,({v^{\prime\prime}_1}^2\,+\,\delta^2) \,+\,
2\,b_{12} \,\delta\,{v^{\prime\prime}_2}^2 \,+\,
2\,b_{13}\,v^{\prime\prime}_1\,v^{\prime\prime}_2\,\delta
\,-\,b_{14}\,({v^{\prime\prime}_1}^2\,+\,3\,\delta^2)\,v^{\prime\prime}_2
  & \nonumber \\
 &
\,-\,b_{24}\,{v^{\prime\prime}_2}^3 \,-\,
b_{34}\,v^{\prime\prime}_1\,{v^{\prime\prime}_2}^2
\,+\,2\,b_{44}\,{v^{\prime\prime}_2}^2 \,\delta & =\;0 \;\;\; .
\label{eq:stan2}
\end{align}
Because we are looking for simultaneous $N_1$ and $N_2$ minima,
these equations have to be solved together with the stationarity
conditions~\eqref{eq:sta} and~\eqref{eq:stacp}. There are two ways
to do this: (a), generate a set of $\{a_i\,,\,b_{jk}\}$ parameters
and use both $N_1$ and $N_2$ stationarity conditions to determine
the 5 vevs; or (b) generate all but 6 of the parameters
$\{a_i\,,\,b_{jk}\}$ and also the 5 vevs and use the stationarity
conditions to determine the remaining 6 potential parameters. We
chose option (b) for two reasons. Firstly, we want one of the minima
to describe the ``real" world. As such, we know what the sum of the
squared vevs should be ((246 GeV)$^2$) at that minimum. It is easier
to use this information by inputing the vevs to begin with.
Secondly, the stationarity
conditions~\eqref{eq:sta},~\eqref{eq:stan2} are cubic in the vevs,
and therefore very difficult to solve analytically. Numerical
methods need to be used, which are difficult to control and
time-consuming. Determining the potential parameters requires
nothing more elaborate than solving a set of linear equations, once
the vevs have been specified. Our method to solve the stationarity
equations, then, consists in: (1) generating values for the vevs,
$\{v_1\,,\,v_2\}$ and
$\{v^{\prime\prime}_1\,,\,v^{\prime\prime}_2\,,\,\delta\}$, and all
but 4 of the $b_{ij}$; (2) using eqs.~\eqref{eq:sta} to determine
the value of the parameters $\{a_1\,,\,a_2\}$; (3) choose the
parameters $\{b_{14}\,,\,b_{24}\,,\,b_{34}\,,\,b_{44}\}$ so that
they satisfy the remaining stationarity equations, namely:
\begin{equation}
\begin{bmatrix} 2\,v^{\prime\prime}_1\,v^{\prime\prime}_2\,\delta &
0 & {v^{\prime\prime}}^2_2\,\delta & 0 \\
\delta\,({v^{\prime\prime}_1}^2\,+\,\delta^2) &
3\,{v^{\prime\prime}}^2_2\,\delta &
2\,v^{\prime\prime}_1\,v^{\prime\prime}_2\,\delta & -
2\,{v^{\prime\prime}}_2\,\delta^2 \\
v^{\prime\prime}_2\,({v^{\prime\prime}_1}^2\,+\,3\,\delta^2) &
{v^{\prime\prime}}^3_2 & v^{\prime\prime}_1\,{v^{\prime\prime}}^2_2
& -2\,\delta\,{v^{\prime\prime}}^2_2 \\
v_1^2 & v_2^2 & v_1\,v_2 & 0 \end{bmatrix} \;
\begin{bmatrix} b_{14}
\\ b_{24} \\ b_{34} \\ b_{44}\end{bmatrix} \;\; =\;\;
\begin{bmatrix} S_1
\\ S_2 \\ S_3 \\ S_4\end{bmatrix} \;\;\; ,
\end{equation}
with
\begin{align}
S_1 & =\;
2\,a_1\,v^{\prime\prime}\,+\,4\,b_{11}\,v^{\prime\prime}\,({v^{\prime\prime}_1}^2\,+\,\delta^2)\,+\,
2\,(b_{12} \,+\,
b_{33})\,v^{\prime\prime}_1\,{v^{\prime\prime}_2}^2\, +\,
b_{13}\,(3\,{v^{\prime\prime}_1}^2\,+\,\delta)\,v^{\prime\prime}_2 \nonumber \\
 & \;\;\;\;\,+\, b_{23}\,{v^{\prime\prime}_2}^3
\nonumber \\
S_2 & =\; 2\,a_2\,v^{\prime\prime}_2 \,+\,
4\,b_{22}\,{v^{\prime\prime}_2}^3 \,+\,
2\,b_{12}\,v^{\prime\prime}_2\,({v^{\prime\prime}_1}^2\,+\,\delta^2)
\,+\, 2\,b_{33}\,{v^{\prime\prime}_1}^2\,v^{\prime\prime}_2 \,+\,
b_{13}\,v^{\prime\prime}_1\,({v^{\prime\prime}_1}^2\,+\,\delta^2)\nonumber \\
 & \;\;\;\;\,+\,3\,b_{23}\,v^{\prime\prime}_1\,{v^{\prime\prime}_2}^2
\nonumber \\
S_3 & =\; 2\,a_1\,\delta \,+\,
4\,b_{11}\,\delta\,({v^{\prime\prime}_1}^2\,+\,\delta^2) \,+\,
2\,b_{12} \,\delta\,{v^{\prime\prime}_2}^2 \,+\,
2\,b_{13}\,v^{\prime\prime}_1\,v^{\prime\prime}_2\,\delta
\nonumber \\
S_4 & =\;  0 \;\;\; . \label{eq:stm}
\end{align}
We chose random values for the potential's parameters, such that the
$b_{ij}$ couplings were of the same order - we considered $b$
parameters in the range $10^{-3}\,\leq\,b_{ij}\,\leq\,10$. To be
certain that the solutions of the stationarity conditions correspond
to minima, we calculated the eigenvalues of the squared scalar mass
matrices and verified that, except for the three zeros corresponding
to the Goldstone bosons, the remaining ones are positive. The mass
matrix expressions for any 2HDM potential may be found in
ref.~\cite{nos}.

\begin{figure}[ht]
\epsfysize=12cm \centerline{\epsfbox{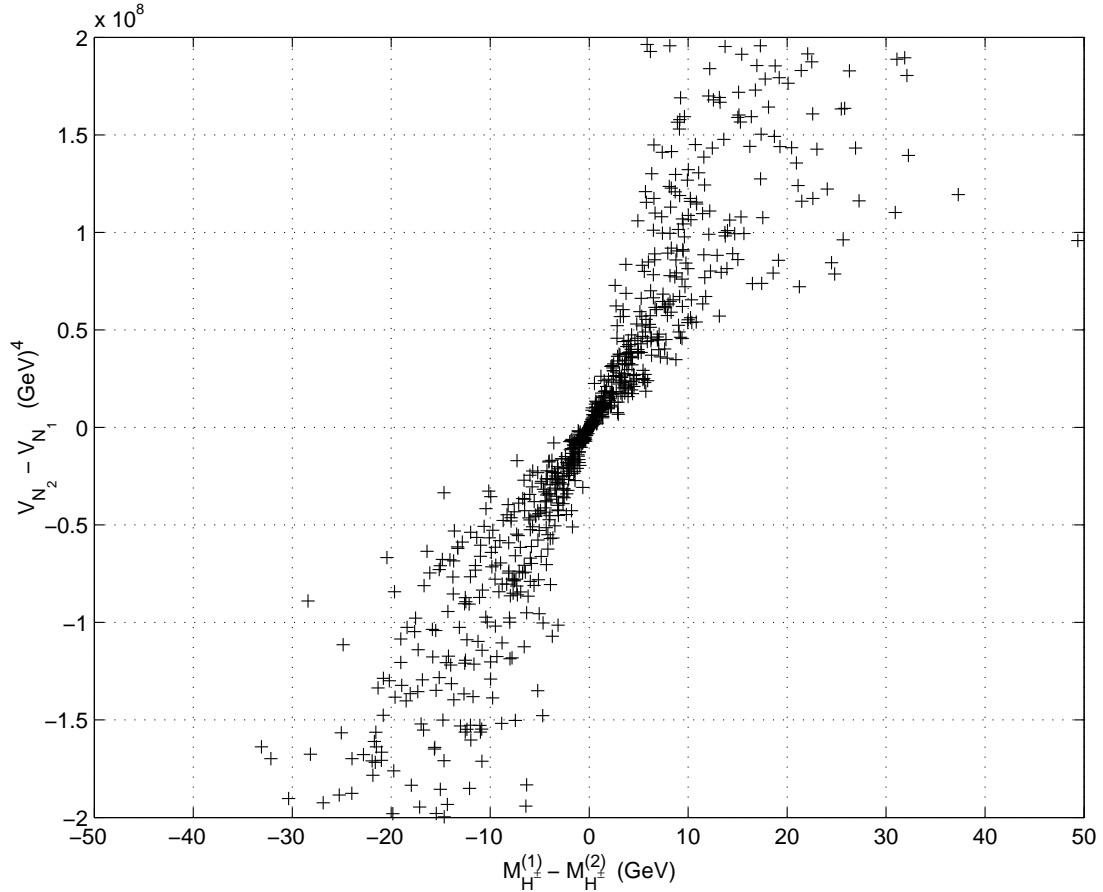}} \caption{Difference
in depth of the potential at two minima, $N_1$ and $N_2$, for the
2HDM model with explicit CP breaking, in terms of the difference of
charged scalar masses.}
\label{fig:pot14}
\end{figure}

Let us see, for instance, if it is possible that the $N_1$ and $N_2$
minima have the {\em same} value for the squared vevs, but {\em
different} scalar masses. In other words, can the 2HDM accommodate
two minima which predict the same $W$, $Z$ and fermion masses, but
different scalar spectra? The answer is, yes. To see this, we solve
eqs.~\eqref{eq:stm} by inputing values for the vevs such that
$v_1^2\,+\,v_2^2\,=\,{v^{\prime\prime}_1}^2\,+\,{v^{\prime\prime}_2}^2
\,+\,\delta^2\,=\,(246\;\mbox{GeV})^2$. We also input the several
$b_{ij}$ parameters and scan the parameter space, accepting only
sets of parameter values for which both $N_1$ and $N_2$ are minima.
To make these minima of some physical interest, we also demanded
that all the scalar masses be larger than 100 GeV, but inferior to 1
TeV. The results we found are illustrated in fig.~\eqref{fig:pot14},
where we plot the difference of potential depths at $N_1$ and $N_2$
against the difference in the charged scalar masses at both minima.
One immediately observes that the $N_1$ and $N_2$ minima are equally
likely to be the deepest minima. We also remark that it is not
difficult to find combinations of $\{a\,,\,b\}$ parameters and vevs
for which these minima coexist: for 10000 sets of parameters for
which $N_1$ is a minimum obeying the criteria described above, about
10\% have a $N_2$ minimum ``alongside". Hence, for the same set of
parameters of the potential we can have two different minima, both
of them predicting the same values for the gauge boson masses and
fermions, but with different spectra of scalar particles. Tunneling
from $N_1$ to $N_2$ would therefore only change the values of Higgs'
masses.

\subsection{Potentials with spontaneous CP breaking}

For the 2HDM where CP is not explicitly broken, all of the couplings
$b_{i4}$ are set to zero, except for $b_{44}$, so that the
stationarity condition of eq.~\eqref{eq:stacp} is trivially
satisfied. Then the number of possible normal minima goes up - we
have a total of {\em nine} possible stationary points from which, as
explained above, we may have a maximum of two non-degenerate minima
away from the origin. For these potentials, though, the stationary
points of the type $N_2$ now correspond to a spontaneous breaking of
the CP symmetry.

For illustrative purposes, let us see if the following situation
might occur: a CP breaking minimum with $v\,=\,246$ GeV above a
normal one~\footnote{Since, we repeat, according to the results of
ref.~\cite{nos} the reverse is not possible.}. At this normal
minimum, the value of $v$ is not fixed {\em a priori}. To achieve
this end we once again solve the stationarity conditions of the 2HDM
potential by generating random values for most of the parameters and
vevs and thus obtaining linear equations on the remaining unknowns.
Solving the linear set of equations thereof resulting produces a
complete set of parameters for the potential, for which we then
proceed to investigate whether both stationary points - the normal
and the CP one - are minima, by analysing the eigenvalues of the
corresponding squared scalar mass matrices.

%
\begin{figure}[ht]
\epsfysize=12cm \centerline{\epsfbox{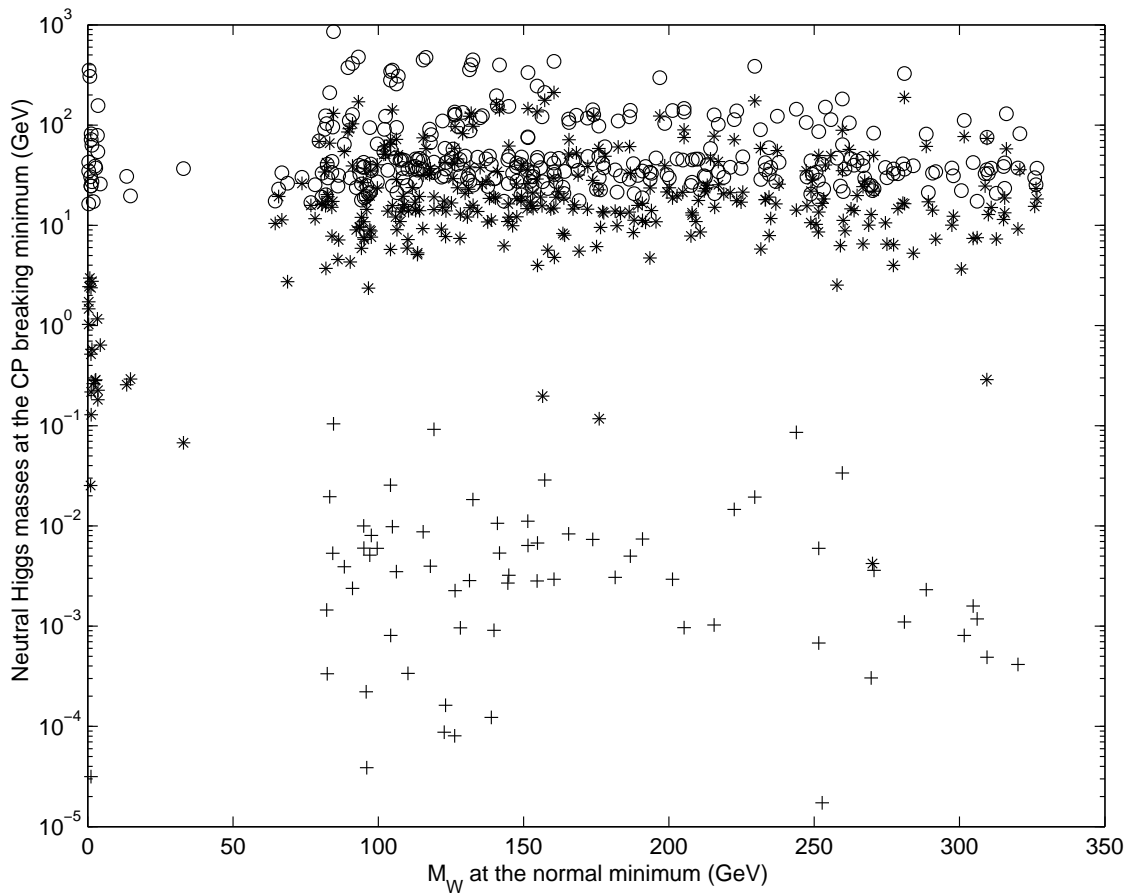}} \caption{Masses of
the neutral Higgs bosons at a CP breaking minimum with $v\,=\,246$
GeV {\em versus} the expected $W$ mass at the deeper normal minimum.
For each combination of potential parameters, the lightest Higgs
mass is represented by crosses, the second lightest mass by stars,
the heaviest one by circles.}
\label{fig:pot10}
\end{figure}
The results we found are shown in fig.~\eqref{fig:pot10}, where we
plot the masses of the neutral Higgs bosons at the CP minimum,
versus the $W$ boson mass expected at the normal minimum. For
visualisation purposes, the vertical axis was limited to values
above $10^{-5}$ GeV, which removed from view several points
corresponding to very light Higgs masses. The normal minimum is
found to have depth similar to that of the CP one, but the resulting
$W$ masses are in general very different. An interesting feature of
the CP violating minimum is also shown in this plot: the neutral
Higgs masses are extremely low, much lower than the most recent
experimental bounds, with the exception of the heaviest neutral
mass, represented by circles in fig.~\eqref{fig:pot10}, which may
reach hundreds of GeV. The second heaviest neutral Higgs (stars in
fig.~\eqref{fig:pot10}) is also very light. For the lightest neutral
Higgs, the maximum value we obtained for its mass was of the order
of 0.1 GeV. One must also emphasize that these sets of parameters
where CP minima coexist with normal ones are extremely difficult to
find. Out of a total of about 10 million normal minima, we only find
CP ones with a sum of squared vevs equal to 246 GeV for some 160 of
those. Our conclusion, then, is that though coexisting normal and CP
minima are numerically possible, they are extremely unlikely, at
least in the range of parameters we chose. When they are found, the
CP minimum seems to be characterized by having a very low mass for
the lightest Higgs boson. This would seem to exclude the strange
possibility that we are ``living" in a CP minimum with standard W
masses, with a normal minimum lying underneath it. Notice, however,
that this does not exclude, at all, the possibility that the
potential has parameters such that the global minimum of the theory
is CP violating. The conclusion of~\cite{nos} only imply that in
this case no normal minimum exists.

We have also considered the situation where the normal minimum has
$v\,=\,246$ GeV, with a CP minimum above it, with vevs determined by
the stationarity conditions. In cosmological terms this would raise
the possibility of a universe ``resting" for some time in such a CP
breaking minimum immediately after the Big Bang, before tunneling
the present normal minimum. Again we found that this situation only
occurs for a very small subset of all the parameter space that was
scanned. The results obtained are very similar to those shown in
fig.~\eqref{fig:pot10}. The $W$ masses found at the CP minimum and
the neutral Higgs masses at the normal minimum tend to be smaller
than 100 GeV. Nevertheless, we have found several cases where the
Higgs masses are compatible with present experimental bounds.

\section{Normal minima in potentials where CP is conserved}
\label{sec:cpc}

In the previous sections we looked at the simultaneous existence of
two neutral minima, one with real vevs, the other with complex ones.
As we showed in section~\ref{sec:vac}, however, it is possible to
have several normal minima with real vevs $\{v_1\,,\,v_2\}$, not all
of them having the same depths. This is already a possibility for
the potentials with soft CP breaking, analysed in the previous
section, but the case we treated there was the most interesting
situation of a CP-breaking minimum coexisting with a normal one.
However, for potentials where CP breaking is impossible the
existence of several normal minima for the same potential is of
interest. As was explained in section~\ref{sec:int}, there are two
different potentials where spontaneous CP breaking is not possible,
depending on the symmetries one has imposed on the 2HDM model. As we
will now show, the normal vacua structure is very different for
these two potentials.

\subsection{Normal vacua in the potential $V_A$}

The stationarity conditions for the potential $V_A$ may be obtained
from the most general ones written in eqs.~\eqref{eq:sta}
and~\eqref{eq:stacp} if one makes $b_{13}\,=\,b_{23}\,=\,0$. Since
for this potential $b_{14}\,=\,b_{24}\,=\,b_{34}\,=\,0$,
eq.~\eqref{eq:stacp} is trivially satisfied and because
$b_{13}\,=\,b_{23}\,=\,0$, eqs.~\eqref{eq:sta} become
\begin{align}
\frac{\partial V}{\partial \varphi_5} & =\;\; 2\,a_1\,v_1 \,+\,
4\,b_{11}\,v_1^3 \,+\, 2\,(b_{12} \,+\, b_{33})\,v_1\,v_2^2 &\;=\,0
\;\;\; \;
\nonumber \\
\frac{\partial V}{\partial \varphi_6} & =\;\; 2\,a_2\,v_2 \,+\,
4\,b_{22}\,v_2^3 \,+\, 2\,(b_{12} \,+\, b_{33})\,v_1^2\,v_2 &\;=\,0
\;\;\; .\label{eq:stava}
\end{align}
We can identify the trivial solution, $v_1\,=\,v_2\,=\,0$, and three
non-trivial ones:
\begin{itemize}
\item Solution I (if $a_2\,<\,0$~\footnote{Requiring that the potential be
bounded from below implies that $b_{11}\,>\,0$ and
$b_{22}\,>\,0$.}):
\begin{equation}
v_1\;=\;0\;\;\; , \;\;\; v_2^2\;=\;-\,\frac{a_2}{2\,b_{22}} \;\;\; .
\end{equation}
At this stationary point the value of the potential is
\begin{equation}
V_I\;=\;-\,\frac{a_2^2}{4\,b_{22}}\;\;\; . \label{eq:VI}
\end{equation}
\item Solution II (if $a_1\,<\,0$):
\begin{equation}
v_1^2\;=\;-\,\frac{a_1}{2\,b_{11}}\;\;\; , \;\;\; v_2\;=\;0 \;\;\; .
\end{equation}
At this stationary point the value of the potential is
\begin{equation}
V_{II}\;=\;-\,\frac{a_1^2}{4\,b_{11}}\;\;\; .\label{eq:VII}
\end{equation}
\item Solution III:
\begin{equation}
v_1^2\;=\;\frac{2\,b_{22}\,a_1\,-\,(b_{12}\,+\,b_{33})\,a_2}{(b_{12}\,+\,b_{33})^2\,-\,4\,b_{11}\,b_{22}}
\;\;\;, \;\;\;
v_2^2\;=\;\frac{2\,b_{11}\,a_2\,-\,(b_{12}\,+\,b_{33})\,a_1}{(b_{12}\,+\,b_{33})^2\,-\,4\,b_{11}\,b_{22}}
\;\;\; .
\end{equation}
At this stationary point the value of the potential is
\begin{equation}
V_{III}\;=\;\frac{b_{22}\,a_1^2\,+\,b_{11}\,a_2^2\,-\,(b_{12}\,+\,b_{33})\,a_1\,a_2}{
4\,b_{11}\,b_{22}\,-\,(b_{12}\,+\,b_{33})^2}\;\;\;.
\end{equation}
\end{itemize}
The solutions I and II correspond to models very similar to the SM
although, for certain realizations of the 2HDM, these solutions may
correspond to cases where either the up or down quarks are massless.
The solution III is the most interesting one, and we will analyse it
in detail. To do so, let us first consider the relations one can
establish between the $\{a\,,\,b\}$ parameters and some of the
physical parameters of the model at a stationary point III: the
neutral CP-even scalar masses $M_h$ and $M_H$, the angle $\beta$
with $\tan\beta\,=\,v_2/v_1$ and the rotation angle $\alpha$ that
diagonalises the matrix of the squared masses of the CP-even
scalars. These relations are found in ref.~\cite{rui}, and are given
by
\begin{align}
a_1 &=\;
-\,\frac{1}{4\,\cos\beta}\,\left[\cos\alpha\,\cos(\beta-\alpha)\,M^2_H\,-\,
\sin\alpha\,\sin(\beta-\alpha)\,M^2_h\right] \nonumber \\
a_2 &=\;
-\,\frac{1}{4\,\sin\beta}\,\left[\sin\alpha\,\cos(\beta-\alpha)\,M^2_H\,+\,
\cos\alpha\,\sin(\beta-\alpha)\,M^2_h\right] \nonumber \\
b_{11}
&=\;\frac{1}{4\,v^2\,\cos^2\beta}\,\left(\cos^2\alpha\,M^2_H\,+
\,\sin^2\alpha\,M^2_h\right) \nonumber \\
b_{22}
&=\;\frac{1}{4\,v^2\,\sin^2\beta}\,\left(\sin^2\alpha\,M^2_H\,+
\,\cos^2\alpha\,M^2_h\right) \;\;\; .\label{eq:rel}
\end{align}
There are more relations, but these four will be all that we will
need for our purposes. Let us assume that we have a set of
parameters of the potential for which the solution III defined above
exists and is a minimum, and for which the solutions I and II exist
as well. The question we now ask ourselves is, under these
circumstances, is it possible that either I or II are also minima,
and deeper than III? Let us start by comparing the value of the
potential at the stationary point I and the minimum III. Making use
of the fact that at any stationary point the value of the potential
is given by $(a_1\,v_1^2\,+\,a_2\,v_2^2)/2$ and using
eqs.~\eqref{eq:rel}, we may rewrite $V_I$ and $V_{III}$ as
\begin{align}
V_I &=\; -\,\frac{v^2}{4}
\,\frac{\left[\sin\alpha\,\cos(\beta-\alpha)\,M^2_H\,+\,
\cos\alpha\,\sin(\beta-\alpha)\,M^2_h\right]^2}{\sin^2\alpha\,M^2_H\,+
\,\cos^2\alpha\,M^2_h} \nonumber \\
V_{III} &=\; -\,\frac{v^2}{4}
\,\left[\cos^2(\beta-\alpha)\,M^2_H\,+\,
\sin^2(\beta-\alpha)\,M^2_h\right] \;\;\; .
\end{align}
Notice that we are rewriting the values of the potential in terms of
the physical parameters (Higgs masses, etc) {\em of the minimum
III}. Can we have $V_I\,<\,V_{III}$? According to the expressions
above that occurs if
\begin{equation}
\frac{\left[\sin\alpha\,\cos(\beta-\alpha)\,M^2_H\,+\,
\cos\alpha\,\sin(\beta-\alpha)\,M^2_h\right]^2}{\sin^2\alpha\,M^2_H\,+
\,\cos^2\alpha\,M^2_h} \;>\;\cos^2(\beta-\alpha)\,M^2_H\,+\,
\sin^2(\beta-\alpha)\,M^2_h \;\; \Leftrightarrow \nonumber
\end{equation}
\begin{align}
\left[\sin\alpha\,\cos(\beta-\alpha)\,M^2_H\right. & \left.\,+\,
\cos\alpha\,\sin(\beta-\alpha)\,M^2_h\right]^2 >\;\;  \nonumber
\\
& >\; \left[\cos^2(\beta-\alpha)\,M^2_H\,+\,
\sin^2(\beta-\alpha)\,M^2_h\right]\,\left(\sin^2\alpha\,M^2_H\,+
\,\cos^2\alpha\,M^2_h\right) \;\;\; ,\label{eq:in}
\end{align}
where the last step was only possible because we are working under
the assumption that the solution III is a minimum - therefore,
$M^2_h\,>\,0$ and $M^2_H\,>\,0$, and the inequality in
eqs.~\eqref{eq:in} is not changed. Developing this inequality leads
to a straightforward conclusion:
\begin{equation}
V_I\,<\,V_{III}\;\;\; \Rightarrow\;\;\; \cos(2\,\beta)\,<\,-\,1
\;\;\;.
\end{equation}
A similar impossibility is found if one investigates the case
$V_{II}\,<\,V_{III}$ when III is a minimum. Which means that, if the
solution III is a minimum, then it is certainly the global minimum
of the theory. Then, the 2HDM potential cannot tunnel from a
phenomenologically acceptable minimum III, where all quarks are
massive, to a deeper solution I or II, where either the up or down
quarks could be massless.

As a curiosity, we may also have a situation where the deepest
minimum is either of the form I or II. In that case, there are two
observations to make: (a) the solution III is necessarily {\em not}
a minimum; and (b), there is the possibility that both solutions I
and II are simultaneously minima. According to eqs.~\eqref{eq:VI}
and~\eqref{eq:VII}, the specific values of the parameters will
determine which of the two solutions corresponds to the deepest
minimum.

\subsection{Normal vacua in the potential $V_B$}

We remind the readers that for the potential $V_B$ we cannot choose
a field basis so that $a_3$ and $\{b_{13}\,,\,b_{23}\}$ are
simultaneously zero. The stationarity conditions for normal vevs in
the potential $V_B$ are very similar to those of eq.~\eqref{eq:sta},
namely
\begin{align}
\frac{\partial V}{\partial \varphi_5} & =\;\; 2\,a_1\,v_1
\,+\,a_3\,v_2\,+\, 4\,b_{11}\,v_1^3 \,+\, 2\,(b_{12} \,+\,
b_{33})\,v_1\,v_2^2 &\;=\,0 \;\;\; \;
\nonumber \\
\frac{\partial V}{\partial \varphi_6} & =\;\; 2\,a_2\,v_2
\,+\,a_3\,v_1\,+\, 4\,b_{22}\,v_2^3 \,+\, 2\,(b_{12} \,+\,
b_{33})\,v_1^2\,v_2 &\;=\,0 \;\;\; . \label{eq:vevb}
\end{align}
The presence of the $a_3$ terms makes it impossible to solve
analytically these equations. We can however follow a similar
strategy to the one we used to treat the most general 2HDM
potential: with polar coordinates $v$ and $\beta$, we obtain an
equation for $v^2$ in terms of $\beta$,
\begin{equation}
v^2\;=\; -\,\frac{1}{\cos^2\beta}\;\frac{2\,a_1\,+\,a_3\,\tan\beta}{
2\,(b_{12} \,+\, b_{33})\,\tan^2\beta \,+\, 4\,b_{11}} \;\;\; ,
\end{equation}
and a quartic equation for $\tan\beta$,
\begin{equation}
a_3\,b_{22}\,\tan^4\beta\,+\,\left[2\,a_1\,b_{22} - a_2\,(b_{12} +
b_{33})\right]\,\tan^3\beta \, +\, \left[a_1\,(b_{12} + b_{33}) -
2\,a_2\,b_{11} \right]\,\tan\beta \, -\, a_3\,b_{11} \;=\;0
\label{eq:tanb}\;\;\; .
\end{equation}

This is an equation very similar to the one we studied in
section~\ref{sec:vac}, so the conclusions we reached there are still
valid: this potential can have, at most, two pairs of non-degenerate
minima away from the origin.

\begin{figure}[ht]
\epsfysize=12cm \centerline{\epsfbox{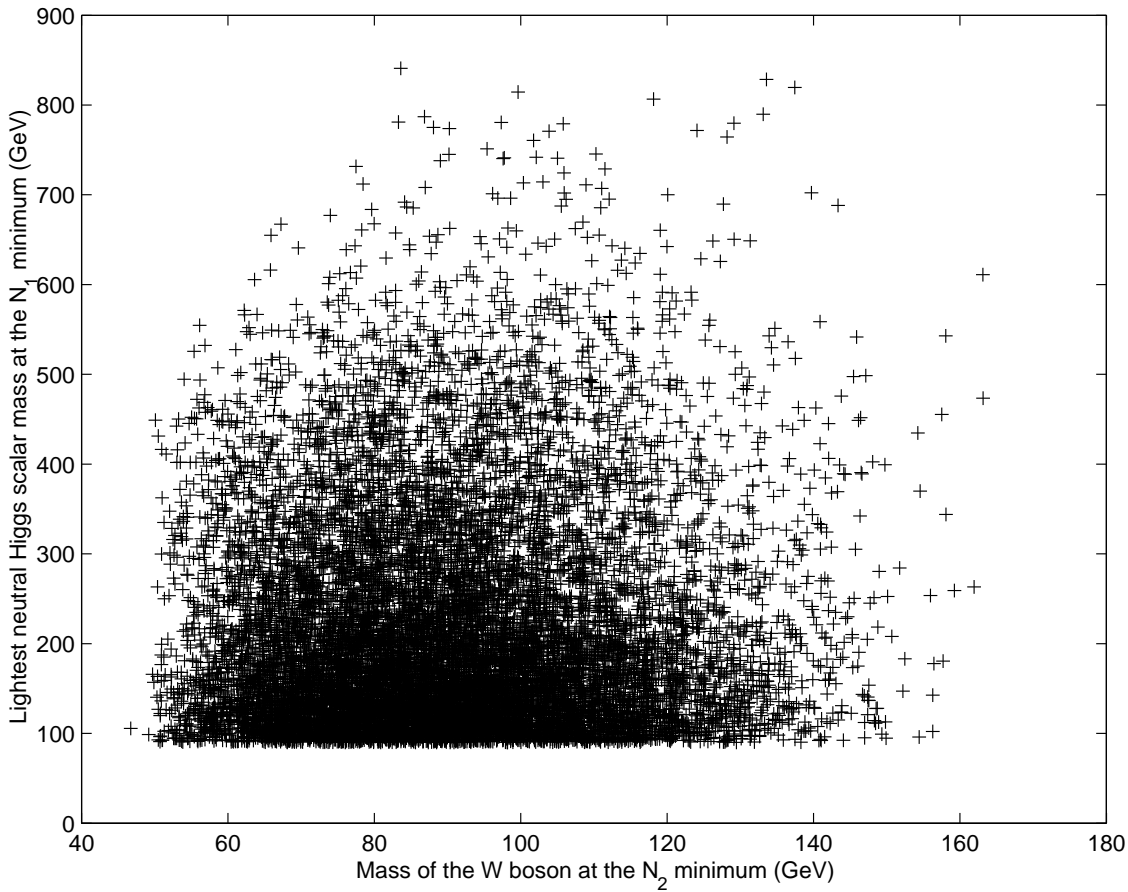}} \caption{Lightest
neutral Higgs scalar mass at the highest $N_1$ minimum versus the
mass of the W gauge boson at the deepest $N_2$ minimum.}
\label{fig:pot7}
\end{figure}

To verify the previous statement, we performed a scan of the
parameter space of the $V_B$ potential searching for such minima. As
before, our procedure was to randomly generate a partial set of
parameters for the potential and two pairs of vevs -
$\{v_1\,,\,v_2\}$, such that
$v_1^2\,+\,v_2^2\,=\,(246\,\mbox{GeV})^2$ for the $N_1$ minimum, and
a second pair $\{\hat{v}_1\,,\,\hat{v}_2\}$, such that
$\hat{v}_1^2\,+\,\hat{v}_2^2$ has a value between (1 GeV)$^2$ and
(1000 GeV)$^2$ for $N_2$. We then use the stationarity
conditions~\eqref{eq:vevb} to determine, solving a set of linear
equations, the parameters $\{a_1\,,\,a_2\,,\,a_3\,,\,b_{33}\}$. So
that minima of the type $N_1$ have some phenomenological relevance
we excluded the combinations of parameters which produced scalar
masses too high (above 1000 GeV) or too low (below about 100 GeV).
We found many such $N_1$ solutions, and for a small subset of those
the minima $N_1$ and $N_2$ coexist. In fig.~\ref{fig:pot7} we show
the plot of the mass of the lightest
neutral Higgs boson mass at the minimum $N_1$ - the standard one,
for which $M_W\,=\,80.5$ GeV - versus the mass expected for the W
boson at the normal minimum $N_2$, {\em when the deepest minimum
found is $N_2$}. For many other points, we found that $N_1$ was the
deepest minimum. The difference in depths of both minima can be
large. The remarkable thing that this figure demonstrates is that we
may be ``living" in a perfectly reasonable $N_1$ minimum, where the
known particle masses are those that have been measured; but
``below" lies a deeper $N_2$ minimum, with exactly the same gauge
symmetries but very different particle mass spectra. In those
circumstances, then, tunneling to the deeper minimum - and to a
universe with very different particles - is, in principle, possible.
We remark that there doesn't seem to be anything particular - any
special combination of parameters of range of values of particle
masses, for instance - characterizing either of the two regimes,
 deeper $N_1$ or deeper $N_2$. We have lightest Higgs masses ranging
from $\sim$ 100 GeV to $\sim$ 800 GeV for the $N_1$ minimum in
either regime.

There are two fulcral observations to make respecting these
conclusions. Firstly, the percentage of parameter space where both
minima were found to coexist is extremely small: out of a generated
15 million $N_1$ minima, a $N_2$ minimum existed in only about
37000, and of those in only about 1/3 of the cases was $N_2$ deeper
than $N_1$. Therefore, this feature of the 2HDM potential is a rare
occurrence, one such case for every thousand trials, but it exists
nevertheless. Secondly, even if the $N_2$ minimum exists and is
below $N_1$, it is not clear whether that means the $N_1$ minimum is
unacceptable or even ``dangerous". If the tunneling time from $N_1$
to $N_2$ is vastly superior to the current age of the universe, then
$N_1$ would be an acceptable vacuum of the theory. The calculation
of tunneling probabilities in models with more than one scalar is a
very complex undertaking~\cite{tun}, and outside the scope of the
present work.

\section{Conclusions}
\label{sec:conc}

We have performed a thorough analysis of the structure of neutral
minima in 2HDM potentials. We have shown that it is possible to have
coexisting neutral minima in such potentials. From a careful study
of the stationarity conditions we were able to establish the maximum
number of possible minima that might exist in those conditions. For
the 2HDM potential with explicit CP breaking, we concluded that
there might only be one normal minimum with real vevs. However, for
this potential, there is no physical distinction between stationary
points with real vevs or with vevs which have a relative complex
phase. We were then able to scan the parameter space of the
potential and discover many combinations of parameters for which
these two types of stationary points are indeed minima, with
different depths. We chose a particulary curious case in which both
of those minima predict identical values for the masses of the known
particles (gauge bosons and fermions) but have completely different
scalar spectra.


In 2HDM potentials with spontaneous CP breaking we found that CP
breaking minima and normal minima can coexist side by side in the
potential, but the combinations of parameters corresponding to this
situation are extremely rare. Further, requiring that the CP minimum
describe the current known particle masses would imply a spectrum of
scalars with very low masses, thus seemingly ruled out by
experiments.

For potentials where CP breaking is not possible, we can still have
normal minima coexisting side by side. For the class of potentials
$V_A$, which have a $Z_2$ symmetry, we found that it is possible to
demonstrate, analytically, that if there is a minimum where both
fields $\Phi$ have non-zero vevs, that minimum is unique and
certainly the global minimum of the theory. However, for another
class of potentials - $V_B$, which have a $U(1)$ symmetry - we may
have combinations of potential parameters for which two normal
minima exist, with different vevs and different scalar spectra. In
particular, we showed that it is possible to have the least deep of
those minima with $v\,=\,246$ GeV, and the deeper one to have a
completely different value for the squared sum of the vevs. This
raises the possibility of tunneling between both minima. In fact,
the results shown in this paper raise several interesting questions
regarding cosmology: is it possible that tunneling between two
normal minima, such as were found for the potential $V_B$, occurs in
less than the age of the universe? Could we have a cosmological
evolution described by a 2HDM potential with spontaneous CP breaking
in which the universe first ``rests" in a CP breaking minimum,
before tunneling or sliding down to the normal minimum that it
currently occupies? And what would be the consequences of such an
evolution in what regards questions of baryogenesis and
matter-antimatter asymmetry? The study of tunneling in models with
several scalars is complex, so these questions lie beyond the scope
of this present work.

\vspace{0.25cm} {\bf Acknowledgments:} We thank L. Sanchez for many
useful discussions. This work is supported by Funda\c{c}\~ao para a
Ci\^encia e Tecnologia under contract POCI/FIS/59741/2004. P.M.F. is
supported by FCT under contract SFRH/BPD/5575/2001. R.S. is
supported by FCT under contract SFRH/BPD/23427/2005.

\appendix
\section{Difference between the depth of the 2HDM potential at two
normal minima}
\label{ap:dem}

As was mentioned in sec.~\ref{sec:vac}, we have the possibility, in
2HDM potentials, of having multiple stationary points with vevs of
the ``normal" type of eq.~\eqref{eq:vevn}. Let us then consider two
such stationary points, one with vevs $\{v_1\,,\,v_2\}$, which we
will call $N_1$, and another, $N_2$, with vevs
$\{\hat{v}_1\,,\,\hat{v}_2\}$. Let us also introduce the notation
used in~\cite{nos}, and define a vector
$X\,=\,[x_1\;\;x_2\;\;x_3\;\;x_4]$, containing the values of the $x$
variables of eq.~\eqref{eq:x} at each of the stationary points. We
also define the quantities $V^\prime_i\,=\,\partial V/\partial x_i$.
For instance,
$V_1\,=\,a_1\,+\,2\,b_{11}\,x_1\,+\,b_{12}\,x_3\,+\,b_{13}\,x_3\,+\,b_{14}\,x_4$.
The vector $V^\prime$ is thus defined as
$V^\prime\,=\,[V^\prime_1\;\;V^\prime_2\;\;V^\prime_3\;\;V^\prime_4]$,
evaluated at each of the stationary points. The stationarity
conditions of the 2HDM potential (eqs.~\eqref{eq:sta}
to~\eqref{eq:tan4} for the most general potential, or the equivalent
ones for the more restricted potentials) imply that, at the $N_1$
stationary point we have
\begin{equation}
X_1\,=\,\begin{bmatrix} v_1^2 \\ v_2^2 \\ v_1\,v_2 \\ 0
\end{bmatrix} \;\;\; , \;\;\; V^\prime\,=\,\left(-\,\frac{V^\prime_3}{2\,v_1\,v_2}\right)_{N_1}\,
\begin{bmatrix} v_2^2 \\ v_1^2 \\ -\,2\,v_1\,v_2 \\ 0 \end{bmatrix}
\;\;\; , \label{eq:xv}
\end{equation}
with analogous definitions at the $N_2$ stationary point for the
vectors $X_2$ and $\hat{V}^\prime$, with the obvious replacements
$v_i\,\rightarrow\,\hat{v}_i$. The quantity
$-\,V^\prime_3/2\,v_1\,v_2$, as was shown in ref.~\cite{nos}, is
related to the value of the squared charged scalar mass. Namely, we
have
\begin{equation}
\left(-\,\frac{V^\prime_3}{2\,v_1\,v_2}\right)_{N_1}\,=\,
\frac{\left(M^2_{H^\pm}\right)_{N_1}}{v_1^2\,+\,v_2^2}\,=\,
\left(\frac{M^2_{H^\pm}}{v^2}\right)_{N_1} \label{eq:vpr}
\end{equation}
with an analogous result for $N_2$. Finally, let us use two more
definitions introduced in ref.~\cite{nos},
\begin{equation}
A\;=\;\begin{bmatrix} a_1 \\ a_2 \\a_3 \\ a_4 \end{bmatrix} \;\;\; ,
\;\;\; B\;=\;\begin{bmatrix} 2 b_{11} & b_{12} & b_{13} & b_{14} \\
b_{12} & 2 b_{22} & b_{23} & b_{24} \\ b_{13} & b_{23} & 2 b_{33} &
b_{34} \\ b_{14} &  b_{24} & b_{34} & 2 b_{44}
\end{bmatrix} \;\; .
\label{eq:ab}
\end{equation}
With this vector and matrix of parameters, we may write the value of
the potential at each of the stationary points in a very concise
manner. As was shown in~\cite{nos} the value of the potential at
$N_1$ is given by
\begin{equation}
V_{N_1}\,=\,\frac{1}{2}\,A^T\,X_1\,=\,-\,\frac{1}{2}\,X_1^T\,B\,X_1
\label{eq:vn1}
\end{equation}
and the vector $V^\prime$ at $N_1$ has a very simple expression:
$V^\prime\,=\,A\,+\,B\,X_1$. Likewise, the value of the potential at
$N_2$ will be given by $V_{N_2}\,=\,\frac{1}{2}\,A^T\,X_2$ and
$V^\prime$, at this stationary point, is given by
$\hat{V}^\prime\,=\,A\,+\,B\,X_2$.

With all necessary definitions introduced we may now demonstrate
eq.~\eqref{eq:difnn}. From the definitions of $X$ and $V^\prime$ and
their values at the stationary points given by eq.~\eqref{eq:xv}, it
is simple to see that we have
\begin{align}
X_1^T\,\hat{V}^\prime
&=\;\;X_1^T\,(A\,+\,B\,X_2)\;\;=\;\;X_1^T\,A\,+\,X_1^T\,B\,X_2\;\;=\;\;2\,V_{N_1}\,+\,X_1^T\,B\,X_2
\nonumber \\
X_2^T\,V^\prime
&=\;\;X_2^T\,(A\,+\,B\,X_1)\;\;=\;\;X_2^T\,A\,+\,X_2^T\,B\,X_1\;\;=\;\;2\,V_{N_2}\,+\,X_2^T\,B\,X_1
\end{align}
where the last equality follows from eq.~\eqref{eq:vn1}, and its
analogue for the $N_2$ stationary point. Notice that, because the
matrix $B$ is symmetric, the two terms $X_1^T\,B\,X_2$ and
$X_2^T\,B\,X_1$ are identical. So that, subtracting these two
equations, we obtain
\begin{equation}
V_{N_2}\,-\,V_{N_1}\;\;=\;\;\frac{1}{2}\left(X_2^T\,V^\prime\,-\,X_1^T\,\hat{V}^\prime\right)
\;\;\; .
\end{equation}
Now, using equations~\eqref{eq:xv} and~\eqref{eq:vpr} we can write
\begin{equation}
X_2^T\,V^\prime\;=\;\begin{bmatrix}\hat{v}_1^2 & \hat{v}_2^2 &
\hat{v}_1\,\hat{v}_2 & 0\end{bmatrix}\;
\left(\frac{M^2_{H^\pm}}{v^2}\right)_{N_1}\;\begin{bmatrix} v_2^2 \\
v_1^2 \\ -\,2\,v_1\,v_2 \\ 0 \end{bmatrix} \;\; = \;\;
\left(\frac{M^2_{H^\pm}}{v^2}\right)_{N_1}\,\left(v_1\,\hat{v}_2
\;-\; v_2\,\hat{v}_1\right)^2
\end{equation}
and an equation entirely analogous to this one for
$X_1^T\,\hat{V}^\prime$. From these two equations one obtains the
result expressed in equation~\eqref{eq:difnn}.

\section{Classification of the several 2HDM potentials}
\label{ap:cla}

As explained in section~\ref{sec:int}, there are many different
types of 2HDM potentials, depending on whether CP is or is not
conserved, and on the types of symmetries that one imposes on the
models. There is however a simple way of grouping those several
potentials in specific categories, characterized by a single number,
which we call the ``index" of the potential. With the definitions of
the real vectors $X$ and $A$ and the matrix $B$ in the previous
appendix, it is trivial to see that the most general 2HDM potential
is written as
\begin{equation}
V\;=\; A^T\,X\,+\,\frac{1}{2}\,X^T\,B\,X \;\;\; .
\end{equation}
Now, $B$ being a real and symmetric matrix, it can be diagonalised
by a given orthogonal transformation $O$, such that
\begin{equation}
O\,B\,O^T\;=\;\begin{bmatrix} \hat{b}_1 & 0 & 0 & 0 \\
0 & \hat{b}_2 & 0 & 0 \\ 0 & 0 & \hat{b}_3 & 0 \\
0 &  0 & 0 & \hat{b}_4
\end{bmatrix} \;\; .
\end{equation}
Accordingly, the vectors $X$ and $A$ are transformed by the matrix
$O$,
\begin{equation}
X\;\rightarrow\,\hat{X}\,=\,O\,X\;\;\; ,\;\;\;
A\;\rightarrow\,\hat{A}\,=\,O\,A\;\;\; ,
\end{equation}
so that the potential is now written as
\begin{equation}
V\;=\;\hat{a}_i\,\hat{x}_i\,+\,\hat{b}_i\,\hat{x}_i^2 \label{eq:vs}
\end{equation}
with a sum on the index $i$ assumed. Because $B$ is a $4\times 4$
matrix and $A$ and $X$ vectors with four elements, we conclude that
through the transformation $O$ we can write the potential in terms
of only {\em eight} quantities $\hat{a}$ and $\hat{b}$ - we say that
this potential has {\em index eight}. Notice that $O$ is {\em not} a
basis transformation, and that we have not reduced the number of
independent parameters of the potential - were we to study the
stationary points of $V$, we would need the original 14 real
parameters of the potential (or 11, with a suitable basis
transformation) to do so. Equation~\eqref{eq:vs} is nothing more
than a simpler way of writing the potential.

For the potential with explicit CP conservation, the matrix $B$ has
zeros on its fourth row and column, except the diagonal element,
$b_{44}$. Also, $a_4\,=\,0$, which means that the last entry of the
vector $A$ is zero. In this case, the transformation $O$ that
diagonalises $B$ has zeros in its fourth row and column, except the
$(4\,,\,4)$ element, which is equal to 1. Then, the rotated vector
$\hat{A}\,=\,O\,A$ {\em still} has a zero in its fourth entry. The
rotated matrix $\hat{B}$ has {\em a priori} four independent
eigenvalues, so this potential ends up being written in terms of
{\em seven} parameters - three $\hat{a}_i$ and four $\hat{b}_i$. The
potential with explicit CP conservation has therefore {\em index
seven}.

What about the potentials $V_A$, $V_B$, $V_C$, for which extra
symmetries have been imposed? Well, for $V_A$ and $V_B$ the matrix
$B$ has further zeros, since $b_{13}\,=\,b_{23}\,=\,0$. The matrix
$O$ is therefore block diagonal, with a $2\times 2$ matrix in its
first two rows and columns and the identity in the third and fourth
positions. The transformation $O$ therefore does not affect the
values of $a_3$, $a_4$, $b_{33}$ and $b_{44}$. The transformed
elements therefore satisfy the conditions $\hat{a}_3\,=\,a_3$,
$\hat{a}_4\,=\,0$ ($a_4$ is already zero for these potentials, since
they explicitly preserve CP), $\hat{b}_3\,=\,2\,b_{33}$ and
$\hat{b}_4\,=\,2\,b_{44}$.

The potential $V_A$ (eq.~\eqref{eq:potA}) has $a_3\,=\,0$ and
$b_{33}\,\neq\,b_{44}$. Therefore, after the transformation $O$, it
will have {\em index six} - two $\hat{a}_i$ and four $\hat{b}_i$
parameters. If one includes a soft breaking term (an $a_3$ term)
then there is an extra parameter - the softly broken $V_A$ has {\em
index seven}. For the potential $V_B$, because $b_{33}\,=\,b_{44}$,
two of the eigenvalues of $B$ are equal. Therefore, there are {\em
three} different $\hat{a}_i$ parameters and {\em three} $\hat{b}_i$
ones - $V_B$ has {\em index six}. Notice that, had we allowed a
massless axion in this model (or in $V_A$) , the index number would
have been reduced by one. Finally, for the potential $V_C$ of
eq.~\eqref{eq:potBr}, the diagonalisation of the matrix $B$ and
corresponding rotation of the vector $A$ leads to the conclusion
that this model has {\em index six}, and that the softly broken
$V_C$ has {\em index seven}. The study of the stationary points of
the potential through the diagonalization of an analogue of matrix
$B$ was done in ref.~\cite{mani}. An interesting observation is that
all the potentials where CP can be broken (explicitly or
spontaneously) have index larger or equal to {\em seven}.

\end{document}